\documentclass 
[journal]{IEEEtran}

\IEEEoverridecommandlockouts
\usepackage{ifpdf}
\usepackage{cite}
\ifCLASSINFOpdf
   \usepackage[pdftex]{graphicx}
 \DeclareGraphicsExtensions{.eps}
\else
   \usepackage[dvips]{graphicx}
  \DeclareGraphicsExtensions{.eps}
\fi
\usepackage[cmex10]{amsmath}
\usepackage{algorithmic}
\usepackage{algorithm} 
\usepackage{algorithmic}
\usepackage{array}
\usepackage{mdwmath}
\usepackage{mdwtab}
\usepackage{eqparbox}
\usepackage{algorithm}
\usepackage{algorithmic}
\usepackage[algo2e]{algorithm2e}

\usepackage{mdwmath}
\usepackage{mdwtab}
\usepackage{amsmath}

\usepackage{amssymb}
\usepackage{dsfont}
\PassOptionsToPackage{bookmarks={false}}{hyperref}
\begin{document}
\title{Packet Relaying Control in Sensing-based Spectrum Sharing Systems}
\author{\IEEEauthorblockN{F.~Foukalas$^{*}$, T.~Khattab$^{*}$, H.V.~Poor$^\dag$} \\
\IEEEauthorblockA{$^*$Electrical Engineering, Qatar University, Doha, Qatar\\
$^\dag$Electrical Engineering, Princeton University, Princeton NJ, USA\\}}

\maketitle

\begin{abstract}
Cognitive relaying has been introduced for opportunistic spectrum access systems by which a secondary node forwards primary packets whenever the primary link faces an outage condition. For spectrum sharing systems, cognitive relaying is parametrized by an interference power constraint level imposed on the transmit power of the secondary user. For sensing-based spectrum sharing, the probability of detection is also involved in packet relaying control. This paper considers the choice of these two parameters so as to maximize the secondary nodes' throughput under certain constraints. The analysis leads to a Markov decision process using dynamic programming approach. The problem is solved using value iteration. Finally, the structural properties of the resulting optimal control are highlighted.    
\end{abstract}


\IEEEpeerreviewmaketitle
\section{Introduction}
Cognitive radio (CR) is a technology whereby the coexistence of licensed users and unlicensed users on the same bandwidth can be achieved. Coexistence can be provided by implementing two different types of schemes known as opportunistic spectrum access (OSA) and spectrum sharing (SS). In OSA CR ~\cite{c1}, a secondary user (SU) i.e., an unlicensed user, is allowed to access the spectrum, which originally allocated to the primary user (PU), only when the spectrum is not used by the PU. In an SS CR ~\cite{c2}, the SU is allowed to transmit simultaneously with the PU in the same spectral band, as long as the interference from the SU does not degrade the performance of the PU to an unacceptable level. Clearly, both types of CR systems guarantee the main design principle of CR which can be expressed as follows: \textit{a secondary system should be "transparent" to the primary system, so as not to interfere with the licensed use of the spectrum.}

\let\thefootnote\relax\footnote{This research was supported by the Qatar National Research Fund under Grant NPRP 08-522-2-211.}

Research on CR technology includes several investigations of OSA and SS systems at both the physical (PHY) and medium access control (MAC) layers. Focusing on the PHY layer, an investigation of achievable capacity and optimal power allocation over fading channels in SS systems is found in ~\cite{c3} while in ~\cite{c4} an analysis of capacity in OSA systems is provided. Moreover, in ~\cite{c5} a new type of CR system called sensing-based SS has been devised and analyzed based on a capacity formulation. On the other hand, investigations dedicated to the MAC layer have focused on issues such as the tradeoff between spectrum sensing and the achievable throughput per frame as studied in ~\cite{c6}, while in ~\cite{c7}, the joint design of the PHY and MAC layers based on the spectrum sensing results has been proposed. Finally, in ~\cite{c8}, stability issues of queues at the MAC layer have been investigated in OSA systems in which the relaying of primary packets can be supported by the secondary system.

In this paper, we examine packet relaying in SS systems that (unlike the OSA system consider in ~\cite{c8}) provide relaying capability of primary packets by the secondary system. This setting is motivated by the fact that the primary network can be served from the secondary network, albeit with competitive behavior due to the sharing nature of the SS system. Specifically, the system is a sensing-based SS; that is, an interference power constraint is imposed for the protection of PU in conjunction with spectrum sensing (SpSe). SpSe introduces more opportunities for simultaneous packet relaying and secondary transmission as pointed out in ~\cite{c5} and ~\cite{c9}. Motivated by this factor, we model and analyze packet relaying in sensing-based SS systems, deriving the achievable primary and secondary throughputs, and showing their interdependencies on SpSe, power control (PoC), the interference power constraint and the utilization factors of the queues. Considering a time slotted system, the sensing results at time $t-1$ will influence the achievable throughput at time $t$ for both PU and SU assuming a specific interference power constraint level and primary utilization factor. To this end, we specify a Markov decision process (MDP) in order to formulate the system's behavior and we proceed with the control of this behavior using a dynamic programming (DP) based solution. Henceforth, we define an immediate reward function that represents the secondary throughput and in the sequel we use value iteration in order to find a solution recursively. For relaxing the complexity of the problem, we investigate several scenarios individually.

The rest of this paper is organized as follows. Section II describes the system model. Section III provides the system analysis, deriving the secondary and primary throughputs. In Section IV, we present the MDP model and Section V describes the value iteration process and the properties of the resulting control. Section VI provides a summary this work and some problems of interest for future work on this topic.

\section{System Model} \label{system}
We assume a sensing-based SS system with one pair of SU-Tx and SU-Rx forming the secondary link and one pair of PU-Tx and PU-Rx forming the primary link with instantaneous channel power gains $g_s$ and $g_p$ respectively. These two links share the same frequency band and thus interference is assumed between them. The instantaneous channel power gains from the SU-Tx to the PU-Rx and from the PU-Tx to the SU-Rx are denoted by $g_{sp}$ and $g_{ps}$ respectively. The considered system is depicted in Fig.1. Both links are assumed to be flat-fading channels with additive white Gaussian noises (AWGNs) denoted by $n_0$ and $n_1$ at the PU-Rx and SU-Rx respectively, where $n_0$ and $n_1$ are assumed to be independent and with the distribution $\mathcal{C} \mathcal{N} (0,N_0)$ (circularly symmetric complex Gaussian). PoC is employed at the SU-Tx in conjunction with spectrum sensing. PoC is deployed according to the rules of sensing-based SS systems whereby the transmit power is allocated with the water-filling algorithm when the channel is idle, denoted by $P_s^{(0)}$ and with an additional interference power constraint protecting the PU ~\cite{c5} when the channel is busy denoted by $P_s^{(1)}$. The SpSe results in four different results known as false alarm and no false alarm under the hypothesis $H_0$ when the channel state is idle with probability $\pi_0$; and detection and missed detection under the hypothesis $H_1$ when the channel state is busy with probability $\pi_1$. The corresponding conditional probabilities of the sensing results are denoted as $P_f$, $(1-P_f)$, and $P_d$, $(1-P_d)$ respectively. SpSe can be accomplished by  an energy detector similar to the one presented in ~\cite{c6} with a sensing time $\tau$ within a frame period $T$ and a sensing threshold $\eta$ compared to an instantaneous sensed signal-to-noise-ratio (SNR) $\gamma_{Se}$.   
  
We also assume that the SU-Tx and the PU-Tx retain their own queues, denoted by $Q_s$ and $Q_p$ respectively, used for forwarding their own packets. In addition, the SU-Tx retains a queue for forwarding packets of the primary network, denoted by $Q_{ps}$. This is known as packet relaying capability by which, if the primary link is in outage condition, then the packets can be forwarded by the secondary node via $Q_{ps}$ thereby helping emptying the queue of the primary user ~\cite{c8}. Instead of the behavior of packet relaying in OSA systems as proposed in ~\cite{c8}, in our case i.e. sensing-based SS, the primary packets in $Q_{p}$ can be forwarded to PU-Rx simultaneously with the secondary packets in $Q_s$ due to the properties of SS systems. Notably, we assume that when the primary packets in $Q_{ps}$ are forwarded, the secondary packets in $Q_s$ are not transmitted.     

\begin{figure}
  \includegraphics[width=95mm,height=70mm]{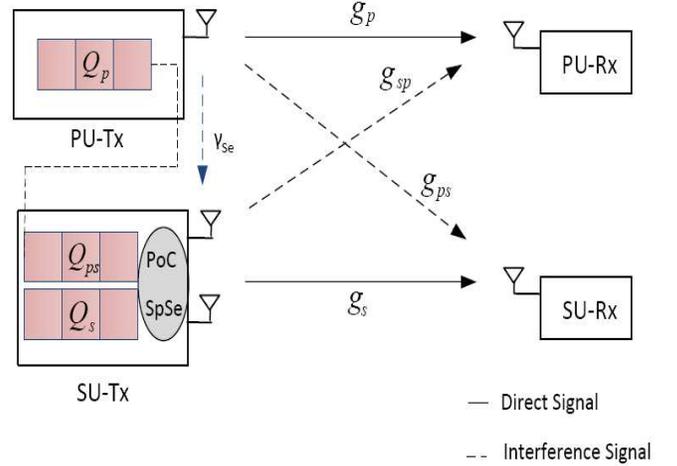}\\
  \caption{System model of a sensing-based SS system with relaying capability.}
  \label{fig:systemmodel}
\end{figure}

Thereafter, the behavior of our system is summarized as follows: 

- the PU-Tx forwards its own packets of $Q_p$, through the primary link with gain $g_p$ as long as the link is not in outage.

- the SU-Tx shares the primary channel used by PU-Tx and forwards its own packets of $Q_s$, through the secondary link with gain $g_s$ as long as the link is not in outage.

- the SU-Tx forwards primary packets of $Q_{ps}$, through the link with gain $g_{sp}$ as long as the primary link is in outage.

We analyze below the system's behavior providing the corresponding formulation. 
 
\section{System Analysis} \label{Analysis} 

In this section, we provide the analysis of secondary and primary throughputs for the system model described above. 

\subsection{Secondary throughput} \label{Sec Thr}

The secondary throughput according to the spectrum sensing results and the transmit power is analysed as follows:

- \textit{No false alarm and unconstrained transmission:}
If the SpSe function results in no false alarms when the PU is idle, then the transmission is accomplished unconstrained with transmit power $P_{s}^{(0)}$ and cut-off level $\beta_s$ with the following probability:

\begin{eqnarray} \label{eq1}
\mu_{s,00} = (1-P_f)e^{(-\frac{\beta_s}{\gamma_s})}\pi_0 
\end{eqnarray}

- \textit{False alarm and constrained transmission:}
If the SpSe function results in false alarms when the PU is not active, then the transmission is accomplished constrained with transmit power $P_{s}^{(1)}$ and cut-off level $\beta_{sp}$ with the following probability: 	

\begin{eqnarray} \label{eq2}
\mu_{s,01} = P_f e^{(-\frac{\beta_{sp}}{\gamma_s})}\pi_0 
\end{eqnarray}

- \textit{Missed detection and unconstrained transmission:} 
If the SpSe function results in missed detection when the PU is active, then the transmission is accomplished unconstrained with transmit power $P_{s}^{(0)}$ and cut-off level $\beta_{s}$ with the following probability: 

\begin{eqnarray} \label{eq3}
\mu_{s,10} = (1-P_d)e^{(-\frac{\beta_{s}}{\gamma_s/(1+\gamma_{ps})})}\pi_1 
\end{eqnarray}

- \textit{Detection and constrained transmission:} 
If the SpSe function results in detection when the PU is active, then the transmission is accomplished constrained with transmit power $P_{s}^{(1)}$ and cut-off level $\beta_{sp}$ defined as follows:

\begin{eqnarray} \label{eq4}
\mu_{s,11} = P_d e^{(-\frac{\beta_{sp}}{\gamma_s/(1+\gamma_{ps})})}\pi_1 
\end{eqnarray}

Finally, since the packet transmission will be done only in situations where there are available packets in the secondary queue, the frame utilization factor becomes $(T-\tau)/T$ due to sending time, and the primary link is not in outage condition with probability $e^{-\beta_p / \gamma_p}$, then the secondary throughput for each of the above cases is obtained as follows: 

\begin{eqnarray} \label{eq5}
\mu_{s} = \frac{T-\tau}{T} \mu_{s,xx}\frac{\lambda_s}{\mu_{s,max}}e^{-\frac{\beta_p}{\gamma_p}}
\end{eqnarray}

where the $xx$ index refers to the aforementioned scenarios.  

\subsection{Primary throughput} \label{Prim Thr}

The primary throughput consists of two different parts, a fixed one related to the conditions of the primary link and its own queue and the other one is the relaying through secondary link which varies according to the sensing results. Thus, the overall primary throughput is defined as follows: 

\begin{eqnarray} \label{eq6}
\nonumber \mu_{p} &=&  e^{-\frac{\beta_p}{(\gamma_p/(1+\gamma_{sp}))}} \pi_1 \\
+ & & \mu_{ps,xx} \left(\frac{\lambda_{ps}}{\mu_{ps,max}}\right)(1-e^{-\frac{\beta_p}{\gamma_p}})
\end{eqnarray}
where the first term represents the throughput of the PU achieved over its own link and thus encompasses the cases in which no outage is experienced on the primary link and the primary queue contains packets, and the second term is related to the current sensing result as long as the $Q_{ps}$ queue contains packets and the primary link experiences outage. As previously for the secondary throughput, the four possible primary relaying throughputs due to the different sensing results are defined below for each case: 

- \textit{No false alarm and unconstrained transmission:}
\begin{eqnarray} \label{eq7}
\mu_{ps,00} = (1-P_f)e^{(-\frac{\beta_s}{\gamma_{sp}})}\pi_0
\end{eqnarray}

- \textit{False alarm and constrained transmission:}
\begin{eqnarray} \label{eq8}
\mu_{ps,01} = P_f e^{(-\frac{\beta_{sp}}{\gamma_sp})}\pi_0
\end{eqnarray}

- \textit{Missed detection and unconstrained transmission:} 
\begin{eqnarray} \label{eq9}
\mu_{ps,10} = (1-P_d)e^{(-\frac{\beta_{s}}{\gamma_{sp}/(1+\gamma_{ps})})}\pi_1
\end{eqnarray}

- \textit{Detection and constrained transmission:} 
\begin{eqnarray} \label{eq10}
\mu_{ps,11} = P_d e^{(-\frac{\beta_{sp}}{\gamma_{sp}/(1+\gamma_{ps})})}\pi_1
\end{eqnarray}

\subsection{Notes}

Notably, we assume in the analysis that the PU's activity $\pi_1$ is equal to the utilization factor  $\lambda_p/\mu_{p,max}$ of the $Q_p$, indicating that the channel is considered active as long as the primary queue $Q_p$ contains packets. Therefore, the primary user's idle probability is obtained as $\pi_0=1-\pi_1=1-\lambda_p/\mu_{p,max}$. In this way, the primary and secondary throughputs are related through the utilization factor of $Q_p$ denoted as $\rho_p$. The utilization factor $\rho_p$ is estimated by the SU-Tx by observing the fraction of idle slots, and measuring the ACK/NACK messages sent by the SU-Rx \cite{c8}. Moreover, it is clear from the above that the utilization factor of the $Q_s$ denoted as $\rho_s$ and the transmit power at secondary link $P_s$ will change the values of the achievable secondary throughput. Thus, we can model the system as a Markov decision process that will include these three parameters (i.e. $\rho_p$, $\rho_s$ and $P_s$) as states and the probability of detection $P_d$ and interference power constraint $I_c$ as actions since they can be adapted by the SU-Tx. Details are given in the next section.    

\section{Markov Decision Process Definition}

In this section, we model the system dynamics as a Markov Decision Process giving first the state and action spaces and finally the dynamic programming problem formulation. 

\subsection{State and action spaces}

The system consists of a finite set of states $S\in(\rho_p,\rho_s,P_s)$ that express the utilization factor $\rho_p$ in the primary queue $Q_p$, the utilization factor $\rho_s$ in the secondary queue $Q_s$ and the transmit power $P_s$ at the SU-Tx. Moreover, the system is controlled through a finite set of actions $A\in(P_d,I_c)$ that includes the probability of detection $P_d$ and the interference power constraint $I_c$. We are interested in the following conditions in each time slot for the system control: 

\begin{eqnarray} \label{eq11}
 (\rho_p,\rho_s,P_s;P_d^{'},I_c^{'})
\end{eqnarray}
that is, the current utilization factor $\rho_p$ in the primary queue $Q_p$, the current utilization factor $\rho_s$ in the secondary queue $Q_s$, the current transmit power $P_s$ at the SU-Tx and the probability of detection $P_d^{'}$ and interference power constraint $I_c^{'}$ used in the previous time slot. The controls applied (i.e. actions that should be taken) in each time slot are the pair of parameters $(P_d,I_{c})$ based on the calculation of the reward function. The reward function in our system is the secondary throughput $\mu_s$. Therefore, the decisions made per slot are: \textit{what should be the probability of detection $P_d$ and the interference power constraint $I_{c}$ level of the SpSe and PoC mechanisms respectively in order to maximize the $\mu_s$}. 

Based on this functionality, the considered system becomes a controlled Markov Decision Process ~\cite{c11} and in this case we can employ a dynamic programming recursion to derive and calculate the optimal control ~\cite{c12} of packet relaying in sensing-based SS systems. 

\subsection{Reward function and transition probability}

First, we define an immediate reward function $g(\rho_p,\rho_s,P_s;P_d^{'},I_c^{'})$ which provides a measure of the maximum secondary throughput $\mu_s$ that can be achieved at the current time slot without consideration of the future throughput given the current utilization factor of the secondary queue $\rho_s$, the current utilization factor of the primary queue $\rho_p$, the current current transmit power $P_s$ and the previous probability of detection $P_d^{'}$ of SpSe and interference power constraint $I_c^{'}$ of PoC. Afterwards, we define the reward function $J(\rho_p,\rho_s,P_s;P_d^{'},I_c^{'})$, which provides a measure of the expected secondary throughput $\mu_s$ going from one state $s$ to another state $s^{'}$ with transition probability

\begin{eqnarray} \label{eq12}
 P_a(s,s^{'}) = Pr(s^{t+1}=s^{'} \mid s_t=s, a_t=a)
\end{eqnarray}
where $a\in A$ is the action and $s\in S$. This is the probability that action $a$ in state $s$ at time $t$ will lead to state $s'$ at time $t+1$. In the defined MDP, assuming the action of SpSe function and the activity of the PU (i.e. the utilization factor) and taking into account the sensing results ~\cite{c5}, the transition probability  $P_a(s,s^{'})$ is obtained as follows:   

\begin{eqnarray} \label{eq13}
&P_a(s,s^{'}) = & \
\begin{cases}
\pi_0 P_f \ , if H_0 \; and \; SpSe \; gives \; H_1 \\ 
\pi_0 (1-P_f)\ , if H_0 \; and \; SpSe \; gives \; H_0 \\
\pi_1 (1-P_d)\ , if H_1 \; and \; SpSe \; gives \; H_0 \\
\pi_1 P_d , if H_1 \; and \; SpSe \; gives \; H_1 \ 
\end{cases}
\end{eqnarray} 

Notably, the probability $P_a(s,s^{'})$ depends on the PU's activity (i.e. the primary utilization factor $\rho_p$) and thus implicitly on the state of the considered MDP and thus the decision making is accomplished recursively.

\subsection{Packet relaying control and costs}

The SU-Tx decides whether to forward the primary packets based on the control of the SpSe and PoC mechanisms changing the probability of detection and the interference power constraint respectively. However, this control comes with overhead costs. We assume the following interpretation as costs for our control approach:

- \textit{Detection control cost: } Increasing the detection probability results in a more costly detection due to the larger number of required samples among other factors. Accordingly a cost factor which is directly proportional to $P_d$ is introduced. We define this cost as follows $\Psi(P_d)=sP_d$ where $s$ is a constant. 

- \textit{Interference control cost: } Reducing the interference power constraint of the PoC stresses the SU-Tx's 	transmission, hence a discount factor is introduced that is increased as $I_c$ is decreased. The concept can be explained as follows: for more protection, less $I_c$ is demanded, and the more the cost for SU-Tx transmission is due to a low power level ~\cite{c14}. We define this cost as $\Phi(I_c)=cP_s^{(1)}$ where $c$ is a constant.    

Below, we define the dynamic programming problem based on the aforementioned characteristics of the defined MDP.

\subsection{Dynamic programming problem formulation}

Having defined the MDP, the system becomes a controlled Markov chain and hence, we can develop a dynamic programming recursion to compute the optimal control as follows:

\begin{eqnarray} \label{eq14}
J(\rho_p,\rho_s,P_s;P_d^{'},I_c^{'}) = \nonumber 
\max_{P_d,I_c}  \lbrace g(\rho_p,\rho_s,P_s;P_d^{'},I_c^{'}) \nonumber \\
-\Psi(P_d) -\Phi(I_c) + \sum_{s^{'}} \beta P_a(s,s{'}) J(\rho_p,\rho_s,P_s;P_d^{'},I_c^{'}) \rbrace
\end{eqnarray}
where the joint optimization $(P_d,I_{pk})$ is performed over the following conditions and selection sets, when the system is in state $(\rho_p,\rho_s,P_s;P_d^{'},I_c^{'})$:

1) the stabilities of all queues are retained i.e. $\lambda_p < \mu_{p,max}$, $\lambda_{ps} < \mu_{ps,max}$ and $\lambda_s < \mu_{s,max}$;

2) the transmit power is constrained over the range $[0,P_{av}]$ through the PoC;  

3) the probability of detection is optimized over a continuous range $P_d\in[0,1]$; and

4) the interference power is optimized over a discrete and bounded power range $[I_{c,min},I_{c,max}]$. 

\section{Value Iteration and Properties of Control}

Solving the DP recursion in (14) results in the optimal control of the probability of detection and the interference power constraint when the system state is $(\rho_p,P_s,\rho_s;P_d^{'},I_c^{'})$. The solution can be obtained using the value iteration method ~\cite{c12}. We confirm the usage of value iteration with the following proposition. 

\textit{Proposition 1:} There exists a stationary optimal control solution of (14) obtainable by value iteration. 

\textit{Sketch of Proof:} The DP terminates when the primary queue $Q_p$ empties, i.e., $\rho_p=0$. The proposed policy is based on the fact that the primary queue is emptied by the factor $\rho_p$ introduced in secondary throughput via $\pi_1$ and thereby the primary queue $Q_p$ will be emptied in finite time. This condition guarantees the existence of a stationary optimal control ~\cite{c12}. The full proof is omitted here for lack of space. 

Using standard value iteration, we start with an initial guess for the maximum reward function and plug it into the DP problem defined in (14). If we consider $k$ iterations, the value iteration problem becomes as follows:

\begin{eqnarray} \label{eq15}
J^{k+1}(\rho_p,\rho_s,P_s;P_d^{'},I_c^{'}) = \nonumber  
\max_{P_d,I_c}  \lbrace g(\rho_p,\rho_s,P_s;P_d^{'},I_c^{'}) \nonumber \\
- \Psi(P_d) -\Phi(I_c) + \sum_{s^{'}} \beta P_a(s,s{'}) J^{k}(\rho_p,\rho_s,P_s;P_d^{'},I_c^{'}) \rbrace
\end{eqnarray}
where based on this formula the following value iteration algorithm is used:
 
 \begin{algorithm}[H]
 \SetAlgoLined
 \KwData{ \\ $\bullet$ Set $P_a(s)$ equal to the transition probability matrix $\forall s\in S$. \\
$\bullet$ For $k=0$, set $J^{k}(s;\alpha)$ and $J^{k+1}(s;\alpha)$ equal to the reward matrix $\forall s\in S$ and $\forall \alpha\in A$.}

 \While{$\max_{s} \mid J^{k+1} - J^{k} \mid < \epsilon$}{
 
  \For{each state $s$ and $\forall \alpha$}{
   calculate $J^{k+1}$ based on \eqref{eq15}\;
   }{
   Set $J^{k}=J^{k+1};$ \\
   Set $k=k+1;$ \;
  }
 }
 \caption{Value Iteration Algorithm}
\end{algorithm}
 
The iteration continues until convergence based on the adopted policy. Once we work out the solution, we store the secondary throughput as a lookup table. The size of the lookup table for each iteration $k$ is $O(\rho_p^k,\rho_s^k,P_s^k,P_d^k,I_c^k)$ where the size depends on the quantized levels of primary and secondary utilization factors, the quantized levels for the channel gain of the secondary link and thus the transmit power, the selected probability of detection and interference power constraint in each iteration. During the system operation, the SU-Tx can determine the optimal secondary throughput from the table according to real-time system parameters.   

To evaluate the performance of the proposed DP control approach, we provide three suboptimal heuristic policies defined as follows \cite{c14}.

\subsection{Variable probability of detection and fixed interference power constraint}

Let us consider the simple case in which the interference power constraint level is fixed and we need to optimize over the probability of detection. In this scenario, the probability of detection $P_d$ is the only control parameter that can be varied to control the packet relaying process;  then, we get the value iteration as follows:

\begin{eqnarray} \label{eq16}
J^{k+1}(\rho_p,\rho_s,P_s;P_d^{'}) = \nonumber 
\max_{P_d}  \lbrace g(\rho_p,\rho_s,P_s;P_d^{'}) \nonumber \\
- \Psi(P_d) + \sum_{s^{'}} \beta P_a(s,s{'}) J^{k} (\rho_p,\rho_s,P_s;P_d^{'}) \rbrace
\end{eqnarray} 

\subsection{Variable interference power constraint and fixed probability of detection}

We now consider the case in which the probability of detection is fixed $P_d=\delta$ and thus the problem is defined as follows: 

\begin{eqnarray} \label{eq17}
J^{k+1}(\rho_p,\rho_s,P_s;I_c^{'}) = \nonumber 
\max_{I_c}  \lbrace g(\rho_p,\rho_s,P_s;I_c^{'}) \nonumber \\
-\Phi(I_c) + \sum_{s^{'}} \beta P_a(s,s{'}) J^{k}(\rho_p,\rho_s,P_s;I_c^{'}) \rbrace
\end{eqnarray}
where the decision is made only upon the interference power constraint $I_c$ state space.

It is interesting to explore the properties of the proposed control. In Fig.2, we plot the secondary throughput $\mu_s$ that represents the derived reward function $J$ from the value iteration method versus the probability of detection $P_d$ for a fixed average power at the secondary link equal to $P_{av}=5dB$, different levels of interference power constraints equal to $I_c=-15dB,I_c=-5dB$ and $I_c=5dB$, and utilization factor at the primary queue to be  $\rho_p=0.1$ and $\rho_p=0.9$. The sensing time is assumed equal to $\tau=0.3ms$ within a frame period of $T=1ms$ and the sensing threshold $\eta$ is implicitly obtained for each $P_d$ assuming a sensed SNR  $\gamma_{Se}=-15dB$. Notably, we do not follow any optimization over sensing time or threshold as has been done in \cite{c6} and \cite{c9}, and thus we choose arbitrary value for the sensing time since our optimization is over $P_d$ and it is not related to the types of optimization presented in these two works. Moreover, the secondary utilization factor is obtained using $\lambda_s=0.5$ and $\mu_{s,max}=0.8$. Upon examining the Fig.2, it is clear that the reward function is maximized for all cases in values of $P_d$ close to $0.7$ and $0.8$. It is also evident that as long as the interference power constraint is relaxed i.e. increased then the value iteration increases as well. Moreover, a high utilization factor $\rho_p$ limits the reward function; however, this effect is negligible in high values of interference power constraints. Notably, the sensing constant is $s=2$ and the number of iterations requires were close to $83$ for both $\rho_p$ values. 

Fig. 3 depicts the results of optimization over the interference power constraint for a fixed probability of detection. We plot again the secondary throughput $\mu_s$ but vs. the interference power values for different probabilities of detection equal to $P_d=0.1$ and $P_d=0.9$ and different utilization factors with equal values as previously. Here, we can see that the control of the interference power constraint does not have any positive result for low probability of detection values, e.g. $P_d=0.1$, and a positive one appears when high probabilities of detection are considered.  This is explained from the fact that the benefit of interference power constraint is introduced in a spectrum sharing system when the detection probability is  high. Notably, the interference control constant is $c=2$ and the number of iterations required were close to $80$ for both $\rho_p$ values. 

\begin{figure}
  \includegraphics[width=95mm,height=70mm]{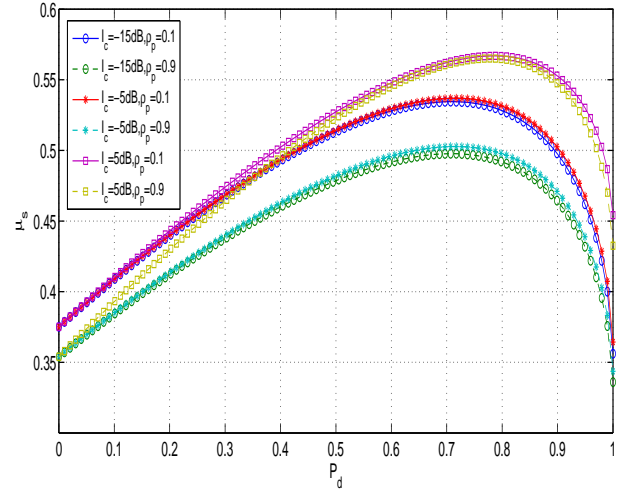}\\
  \caption{Secondary throughput $\mu_s$ vs. probability of detection $P_d$, for average transmit power at the SU-Tx $P_{av}=5dB$, for different levels of interference power constraints $I_c=-15dB,-5dB,5dB$ and utilization factors $\rho_p=0,1.0.9$.}
  \label{fig:2}
\end{figure}

\begin{figure}
  \includegraphics[width=95mm,height=70mm]{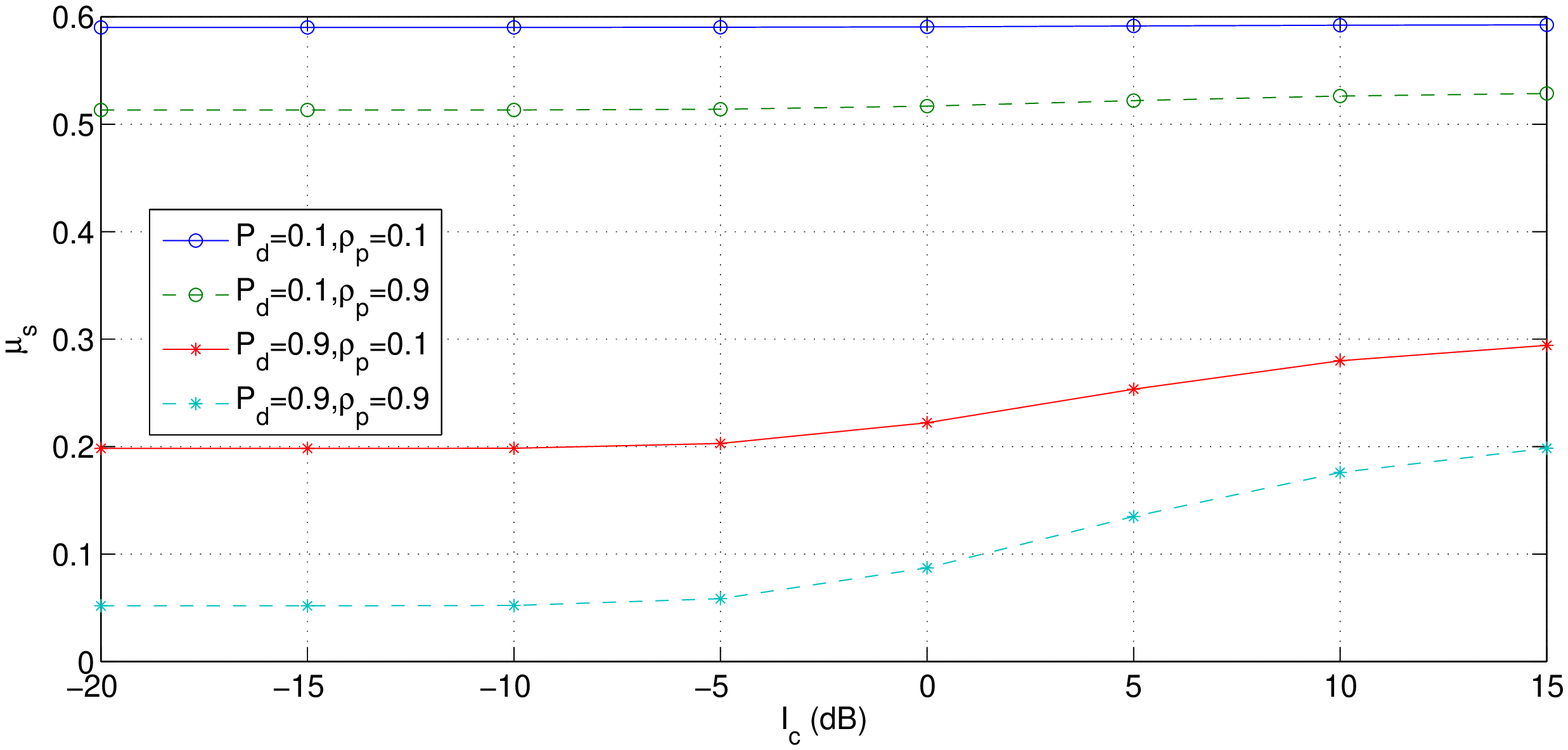}\\
  \caption{Secondary throughput $\mu_s$ vs. interference power constraint $I_c$, for average transmit power at the SU-Tx $P_{av}=5dB$, for different probabilities of detection $P_d=0.1,0.9$ and utilization factors $\rho_p=0.1,0.9$.}
  \label{fig:3}
\end{figure}

\subsection{Joint control: variable probability of detection and variable interference power constraint}

In the case of joint control of the probability of detection and interference power constraint, the value iteration DP problem is formulated now as follows:

\begin{eqnarray} \label{eq17}
J^{k+1}(\rho_p,\rho_s,P_s;P_d^{'},I_c^{'}) = \nonumber 
\max_{P_d,I_c}  \lbrace g(\rho_p,\rho_s,P_s;P_d^{'}) \nonumber \\
- \Psi(P_d) -\Phi(I_c) + \sum_{s^{'}} \beta P_a(s,s{'}) J^{k} (\rho_p,\rho_s,P_s;P_d^{'}) \rbrace
\end{eqnarray}

This implementation differs from the previous one since now the cost-to-go function, i.e. the reward minus the costs, is controlled by the joint leverage of the probability of detection and interference constraint. 

Fig. 4 shows the values of control parameters, i.e. $I_c$ and $P_d$, which maximize the secondary throughput for a specific average transmit power $P_{av}$ at the secondary link when the joint control is accomplished. We see that for high power regions i.e. $P_{av}>10dB$ the probability of detection takes the value $P_d=0.7$ for each interference power constraint value $I_c$. For low average regions, i.e. $P_{av}<-5dB$, the maximization is achieved at $P_d=0$ for interference constrain equal to $I_c<-10dB$ and at $P_d=1$ for interference constrain equal to $I_c>-10dB$ highlighting the important role of interference power constraint at the low power region. These conclusions can be confirmed by Fig.2 and Fig.3 respectively in which the separate control has been presented.      

\begin{figure}
  \includegraphics[width=95mm,height=70mm]{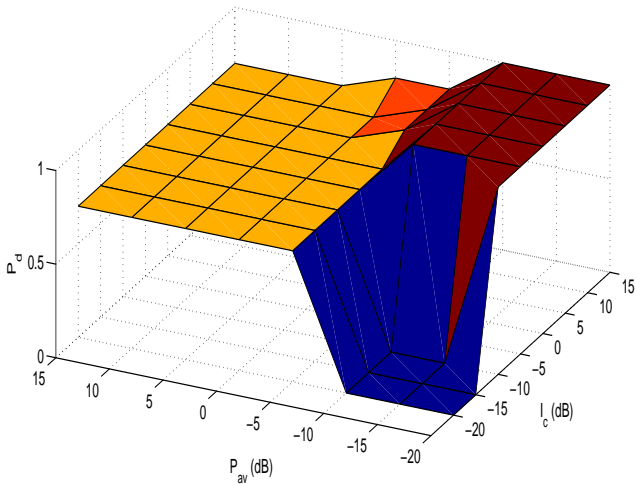}\\
  \caption{Values of control parameters $P_d$ and $I_c$ for a specific range of the average power $P_{av}$ at the secondary link that maximize the secondary throughput.}
  \label{fig:4}
\end{figure}

\section{Summary and Future Work}

In this paper, we have investigated the problem of packet relaying control in sensing-based spectrum sharing systems. We have analyzed the system highlighting the interdependencies between the system components, and we have defined a Markov decision process for modelling the system behavior, which was divided into a finite state space and a finite action space. The model resulted in a dynamic programming problem that was solved using the value iteration. The objective was the maximization of the secondary throughput by managing the spectrum sensing and the interference power constraint. Hence, we have designed and evaluated practical heuristics achieving optimal performance by first separating these two control parameters and then combining them.

An interesting issue for further study based on the above analysis and results is policy improvement using policy iteration finding exact policies under certain conditions and reinforcement learning for estimating the optimal action-state function without requiring a model of the environment. This approach should allow manipulation of the policies directly, rather than finding them indirectly via the optimal value function. Moreover, the cost for particular policies will play an additional role towards the depiction of optimal policies with the lowest costs. Finally, we will be able to use simulation to compare several optimal policies highlighting the performance capabilities of the proposed control in packet relaying.

\end{document}